\documentclass[11pt, twocolumn]{article}
\AtBeginDocument{%
  }

\usepackage{geometry}
\usepackage{graphicx}
\usepackage{listings}
\usepackage{array}
\usepackage{dblfloatfix}
\usepackage{xcolor}
\usepackage{hyperref}

\begin{document}

\title{AgentSLA : Towards a Service Level Agreement for AI Agents}

\author{Gwendal Jouneaux\\\href{mailto:gwendal.jouneaux@list.lu}{gwendal.jouneaux@list.lu}\\Luxembourg Institute of\\Science and Technology \and Jordi Cabot\\\href{mailto:jordi.cabot@list.lu}{jordi.cabot@list.lu}\\Luxembourg Institute of\\Science and Technology}

\date{}

\maketitle

\begin{abstract}
    AI components are increasingly becoming a key element of all types of software systems to enhance their functionality.
    These AI components are often implemented as AI Agents, offering more autonomy than a plain integration of Large Language Models (LLMs), moving from a  Model-as-a-Service paradigm to an Agent-as-a-Service one, bringing new challenges to the development of smart software systems.
    Indeed,  while support for the design, implementation, and deployment of those agents exist, the specification of Quality of Service (QoS) and definition of Service Level Agreements (SLAs) aspects for those agents, important to ensure the quality of the resulting systems, remains an open challenge.
    Part of this is due to the difficulty to clearly define quality in the context of AI components, resulting in a lack of consensus on how to best approach Quality Assurance (QA) for these types of systems.
    To address this challenge, this paper proposes both a quality model for AI agents based on the ISO/IEC 25010 standard, and a domain specific language to support the definition of SLAs for the services provided by these AI agents.
\end{abstract}

\section{Introduction}
\label{sec:intro}

In the last years, we have observed an exponential increase in development effort and integration of AI models in software systems, in particular Machine Learning (ML) models and Large Language Models (LLMs).
The integration of these models enabled the development of  AI-enhanced software systems including smart components such as chatbots, image generators or recommended systems.
These models were typically integrated to standard software using the Model-as-a-Service paradigm~\cite{gan2023model} where some software features were implemented as prompts sent to such LLMs.

With the current increased interest on Agentic AI and AI Agents, this trend is evolving to a new Agent-as-a-Service paradigm where the integration is performed via a collaboration with more autonomous agents able to perform more complex tasks.  These agents can either be deployed on the cloud or locally and will be connected to traditional software using emerging protocols such as Google Agent2Agent Protocol (A2A)\footnote{https://a2aprotocol.ai/}.

As for any Service-Oriented Architecture, the agent consumer generally expects properties such as performance, availability or fault tolerance to be good enough for its use.
This expected Quality of Service (QoS) can be formalized and potentially enforced through the use of Service Level Agreements (SLAs). SLAs allow service consumers to explicit the minimum quality expected from a service provider to function properly.
This quality of service agreement must be acknowledged by both the service consumer and provider.
SLAs are typically composed of a set of conditions called Service Level Objectives (SLOs) defining a threshold for a given metric.
For instance an SLO can take a form such as: \textit{the response time should be less than 50 milliseconds}.
Several formal languages to specify and automatically process SLAs have been proposed.

The advantage of automatically processing SLAs is the possibility to generate related artifacts, such as monitors to dynamically evaluate the required metrics.
At the core of SLAs is a quality model defining the quality characteristics of the service at stake and the relevant metrics associated to them.
Among the multiple existing quality models, the current standard for software products is the ISO/IEC 25010.
This standard defines a quality model composed of nine main characteristics which are further subdivided into subcharacteristics.

Nevertheless, currently there is no formalism to support the definition of SLAs for AI agents.
One of the missing elements for the establishment of such formalism is the lack of a clear and agreed quality model for AI Agents. Although quality models for AI software exist, there is no quality model dedicated to AI agents.
Furthermore, while a decent part of these quality models overlap, there is no consensus on a good quality model for AI software.
This is mainly due to the fact that quality is a notoriously difficult aspect to pinpoint~\cite{neil2017software}.

To address these challenges, this paper proposes both a domain-specific language (DSL) called AgentSLA to specify Service Level Agreements for AI Agents and the associated quality model, extending the ISO/IEC 25010 standard with quality characteristics dedicated to AI Agents.
In addition, we provide a Python implementation of AgentSLA in the form of a validating parser.

The rest of the paper is structured as follows: Section~\ref{sec:background} reviews related work on quality models for AI software and SLA languages for AI agents, Section~\ref{sec:quality} details the additional quality characteristics for AI software, our extension of the ISO 25010 standard to address them, and the core set of metrics related to these characteristics, Section~\ref{sec:dsl} presents the proposed SLA language for AI Agents, and finally Section~\ref{sec:discussion} and Section~\ref{sec:conclusion} discuss and conclude this paper.

\section{Related Work}
\label{sec:background}

In this section, we discuss previous work related to the concept of SLA for AI agents.
In particular, we present existing work on quality models specifically designed for AI software, languages to specify SLAs, and SLA-awareness in the context of AI agents.

\subsection{Quality Models of AI Software}

With the growing importance of AI-based systems, the quality of those systems has become equally important.
The ability to ensure software quality and provide quality assurances has been identified as a challenge for current AI software research~\cite{felderer2021quality}.
Yet, the quality model on which these assurances are based is still  evolving.

In the past years, the research community proposed multiple quality models~\cite{gezici2022systematic,ali2022systematic}.
Some of those models are created from the ground-up, while other map their quality attributes to the ones of ISO/IEC 25010~\cite{siebert2020towards,nakamichi2020requirements,poth2020quality,kuwajima2019adapting,kuwajima2020engineering,pons2019priority}.

In our approach, we consider AI agents as software with an AI component, hence we focus on the quality models mapped to the ISO standard for software quality. 
However, most of those models do not present a full picture of AI software quality.
Among the six papers mapping to ISO 25010, only two discuss sustainability as a relevant quality aspect~\cite{pons2019priority,siebert2020towards} and two overlook explainability/accountability~\cite{kuwajima2020engineering,poth2020quality}.
Furthermore, when considering AI agents, the autonomy level of the agent affects the interaction with the user increasing or decreasing the quality of the interaction.
This aspect of quality is currently not discussed in any quality model.

Beyond the characteristics covered by quality models, other formalisms also structure and report metrics about machine learning models or related artifacts. For AI models, Mitchell \textit{et al.} proposed Model Cards~\cite{mitchell2019model} used for model reporting.
Model Cards describe the AI model in terms of intended use, data, performances and ethical considerations, among other things.
Extensions of this work include HuggingFace implementation adding sustainability data~\cite{ModelCardsCO2} (\textit{e.g.,} cloud provider location, training time, hardware, and estimated carbon emissions) and the recent Sustainability Model Card~\cite{jouneaux2025sustainability} aiming at reporting the AI models sustainability impact using a dedicated formalism allowing automatic analysis.
For datasets, Dataset Cards~\cite{DatasetCards} allows defining standard information such as provenance, authorship, license, and tasks for which the dataset is suitable.
DescribeML~\cite{giner2023domain} additionally describes social concerns potentially leading to bias and provides a dedicated language and tool support for its specification.
Finally, Croissant~\cite{akhtar2024croissant} uses a JSON notation that is both human-readable and compatible with existing tools and frameworks, providing additional interoperability, portability, and discoverability to datasets. All these characteristics could be used to inform a new quality model for AI agents.

\subsection{SLA Specification Languages}

Quality models are especially useful in the context of a Service Level Agreement that imposes a contract between a provider and a client with respect to specific quality values to be respected. 

Various languages have been proposed in the literature to specify SLAs for web-services.
Among the most notable of them, WSLA~\cite{WSLA}, SLAng~\cite{SLANG}, SLA*~\cite{SLA-star}, and WS-Agreement~\cite{WS-Agreement} were designed for synchronous services, while the more recent AsyncSLA~\cite{Oriol2024AsyncSLA} targets  asynchronous services. While these formalisms allow the specification of SLAs for any service, more tailored languages have been developed to address different types of services or domains such as cloud computing~\cite{maarouf2015review,qazi2024service,wazir2016service,mustafa2019analyzing,ezzeddine2021design}, security~\cite{nicolazzo2024service,casola2013specs,d2015towards}, Internet of Things~\cite{alqahtani2019service,SLA-IoT,WIoT-SLA}, or optical networking~\cite{fawaz2004service}.

However, none of these approaches provides dedicated abstraction for the definition of SLAs, and in particular Service-Level Objectives (SLOs) defining the QoS expectations, for AI agents.
In particular, the contextual nature of agents can lead to drift in the quality of service and uncertainty in the metrics evaluation, which are not accounted for in the existing languages.

\subsection{SLA-Awareness and AI Agents}

Even if not fully formalized, recent works providing some SLA-awareness for specific types of AI agents exist.
For instance, Thangarajah \textit{et al.}~\cite{thangarajah2025slaawareness} proposed an SLA-aware task orchestrator for AI coding assistant.
In this work, they present the idea that Code Large Language Models used for coding assistant should have different non-functional requirement (for latency in this case) depending on the coding task.
Tasks such as code generation and code completion should prioritize Time-To-First-Token for improved iteration process and real-time edition, respectively.
On the other hand, code refactoring and code summarization should prioritize having a low end-to-end latency, as the user would typically wait for the complete output.

We aim to provide a fully formalized language for SLAs for all types of agents. 

\section{Quality Model for AI Agents}
\label{sec:quality}

In the current landscape of AI software quality, there is no consensus on a general quality model for AI agents. Proposing such quality model is the goal of this section.

Obviously, we are not going to start from scratch. Some quality characteristics are  included in existing quality models~\cite{ali2022systematic} and could be directly reused: 
\textbf{Correctness} indicate that the AI software produce the expected result, encompassing accuracy, precision and recall of the model.
\textbf{Model Relevance} represent the adequacy of the model architecture to the data and problem to solve.
\textbf{Efficiency} describe the time behavior of the AI software, such as Time-To-First-Token or end-to-end response time.
\textbf{Robustness} denote the resilience to perturbations and is often associated to security concerns.
\textbf{Fairness} identify the different types of bias (e.g., racism, sexism) of the AI software.
Finally, \textbf{Interpretability} characterize the ability to understand the decision process.

\begin{table}[!htb]
\centering
\small
\begin{tabular}{|ll|}
\hline
Correctness                                               & Sustainability      \\
-- \textit{Accuracy}                                               & Interoperability    \\
-- \textit{Precision}                                              & Autonomy            \\
-- \textit{Recall}                                                 & Understandability   \\
Model relevance                                           & -- \textit{Interpretability} \\
Robustness                                                & -- \textit{Transparency}     \\
-- \textit{Reliability}                                            & -- \textit{Explainability}   \\
-- \textit{Security}                                               & -- \textit{Accountability}   \\
Efficiency                                                & Output properties   \\
-- \textit{Time-To-First-Token}~\cite{thangarajah2025slaawareness} & -- \textit{Conciseness}      \\
-- \textit{E2E response time}~\cite{thangarajah2025slaawareness}   & -- \textit{Consistency}      \\
-- \textit{Training} time                                          & -- \textit{Creativity}       \\
Fairness                                                  & -- \textit{Diversity}        \\ \hline
\end{tabular}%
\caption{AI software relevant quality characteristics}
\label{tab:chacteristics}

\end{table}

For our quality model, we extend this set with additional characteristics that should also be accounted for.   Table~\ref{tab:chacteristics} summarizes all the relevant quality characteristics.

\begin{figure*}[!b]
    \centering
    \includegraphics[width=0.96\textwidth]{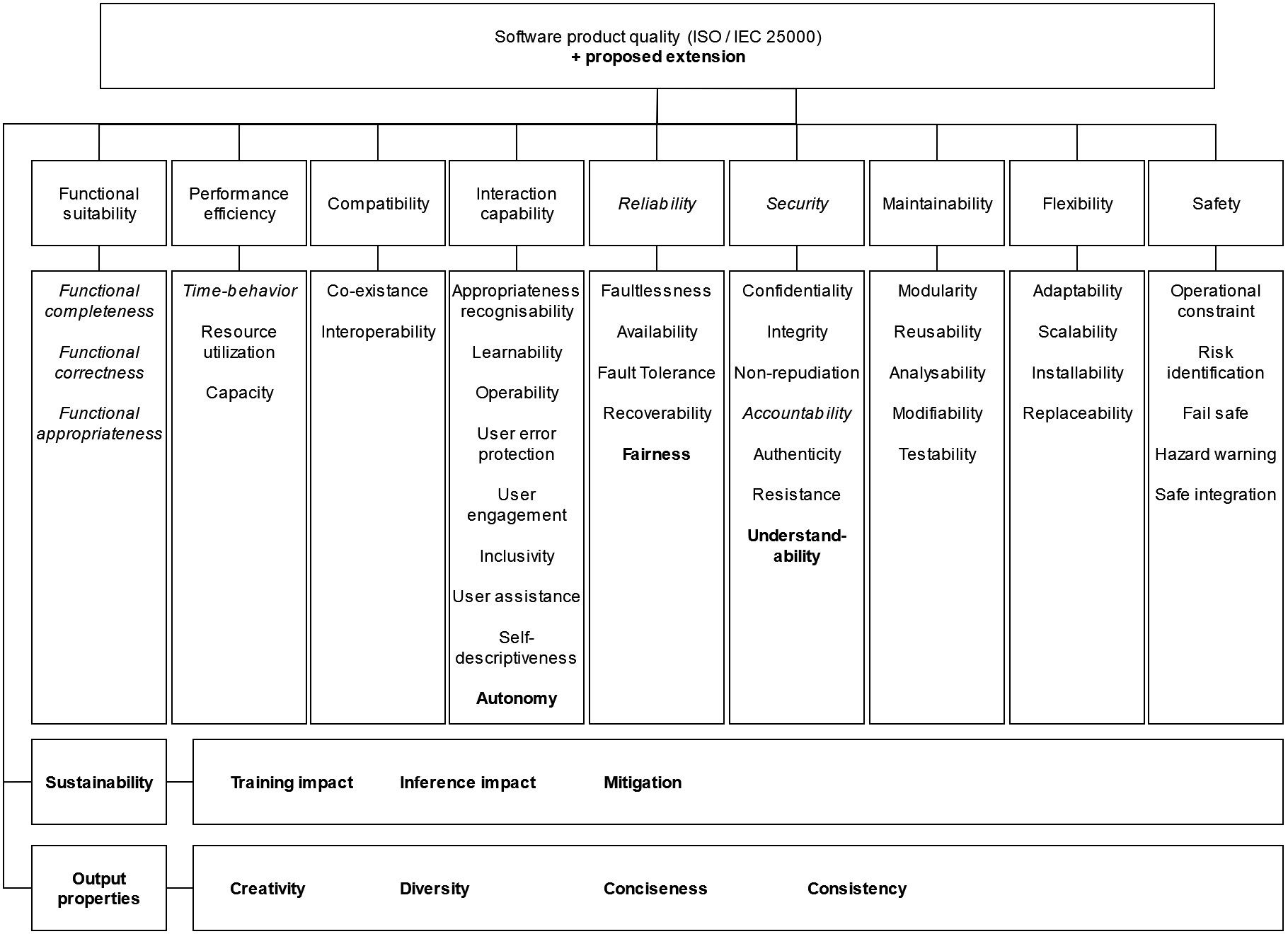}
    \caption{ISO/IEC 25010 quality model for software systems with the proposed extension for AI software system}
    \label{fig:ISO}
\end{figure*}

This includes new dimensions we consider important such as Sustainability.
Indeed, the rising interest for sustainable AI, also known as Green AI~\cite{verdecchia2023systematic}, shows that sustainability is also one of the main extra-functional properties of AI software to be addressed. There is no yet an existing standard for consistently collecting sustainable metrics ~\cite{cruz2025greeningaienabledsystemssoftware}.
Thus, we based our quality characteristics and metrics on the Sustainability Model Card~\cite{jouneaux2025sustainability}, adding a \textbf{Sustainability} characteristics in the quality model addressing the training impact, inference impact, and mitigation for these impacts tackled by the model or platform provider.  

We also add the \textbf{Autonomy} characteristic.
In the current trends in AI research, Agentic AI and AI Agent become more and more prevalent. 
This new paradigm implies a thought-process from the agent and the ability to refine answers autonomously.
This refinement can require new information from the user, creating different levels of autonomy depending on its frequency.

Furthermore, AI agents are now more complex and can be composed of multiple sub-components, introducing the need for \textbf{Interoperability}.
These components can be tools connected using protocols such as the Model Context Protocol (MCP), or multi-agent system delegating part of the task to other agents.

In this context, understanding the decision process is critical to assess the quality of the result.
This is why properties like transparency, explainability or accountability have been introduced in some quality models~\cite{poth2020quality,siebert2020towards,pons2019priority,kuwajima2019adapting}.
We regroup these aspects under the \textbf{Understandability} characteristic.

Finally, the goal of the majority of the AI software is to provide an output to the user in the form of generated text, code or image.
The quality of these outputs can change significantly while remaining correct with respect to the provided prompt.
To tackle these concerns, we propose to add anew \textbf{Output properties} characteristics to our quality model.
This dimension encompasses extra-functional properties of the generated output such as the conciseness of the answer, the consistency among outputs for the same prompt, the creativity of the model, and the diversity of possible output.

\pagebreak
As AI software is still software, we decided to formalize our quality model as an extension of the ISO/IEC 25010 standard for software product quality.
Figure~\ref{fig:ISO} shows the existing standard and our proposed extension, denoted by bold text when something was added and by italicized text when something was extended.
The existing model is structured into nine sections, each composed of multiple quality characteristics.

In addition to the existing sections, we added two new sections addressing the sustainability aspect and output properties of the AI software, regrouping the related characteristics previously detailed (e.g., training impact for the former and conciseness for the later).
The remaining characteristics were either added to an appropriate existing section or extended an existing characteristics' definition to encompass the additional concern.
The \textbf{Autonomy} characteristic representing the level of User/Agent interaction required,  was included in the \textit{Interaction capability} section.
\textbf{Fairness} represents the \textit{Reliability} of the agent concerning ethical biases.
In the case of \textbf{Understandability}, the accountability characteristic already present in the \textit{Security} section was extended, while the rest of the sub-characteristic remains in the new \textbf{Understandability} characteristic.
The extension of the accountability characteristic takes the form of a new definition of the term to address traceability of entities actions through the AI model.
Other extensions include \textit{Functional completeness}, \textit{Functional correctness} and \textit{Functional appropriateness} broadening the scope to completeness of the model   capabilities, accuracy of the result, and model relevance respectively.
\textit{Time-behavior} is extended to encompass the model training time and Time-To-First-Token emerging from the iterative nature of agents.
The \textit{Reliability} and \textit{Security} sections' extensions address the robustness aspect of AI models.
Finally, the existing interoperability characteristic is left untouched, as its definition remains the same.

To complement this quality model, we provide a set of core metrics that can be used to assess the proposed characteristics.
These are detailed in Table~\ref{tab:metrics} available in the appendices.
This table lists the name, related characteristic in the quality model, and definition of all the proposed metrics.
This list is neither complete nor exhaustive as model-specific metrics (e.g., depth of the decision tree) and use-case specific metrics (e.g., ROUGE, BLEU) are not included as part of this core set.
Furthermore, some of the characteristics defined in the quality model, such as creativity and diversity, do not have any precise known metrics yet.

\section{AgentSLA: a DSL to define AI-Agent SLAs}
\label{sec:dsl}

In this section, we propose a domain-specific language (DSL) to support the definition of SLAs based on the previously detailed quality model. This DSL facilitates the creation of new SLAs for AI Agents and their automatic processing in any type of discussion, selection or implementation of AI agents as part of a software system.  We call this DSL AgentSLA.

A DSL is a language specially designed to model software for a specific domain (network, kernel configuration, or,  as in this case, SLA). A DSL includes two main components: the abstract syntax and the concrete syntax.
The abstract syntax describes the structure of the language and the way different language primitives can be combined, independently of any particular representation.
The concrete syntax describes one or more notations for the language, covering textual or graphical representation of the language primitives.
In the case of AgentSLA, this concrete syntax is based on JSON to facilitate its integration to existing AI agent protocols such as A2A\footnote{Agent2Agent Protocol JSON Specification: \url{https://github.com/a2aproject/A2A/tree/main/specification/json}}. 

\subsection{AgentSLA Abstract Syntax}

\begin{figure*}[!htb]
    \centering
    \includegraphics[width=\textwidth]{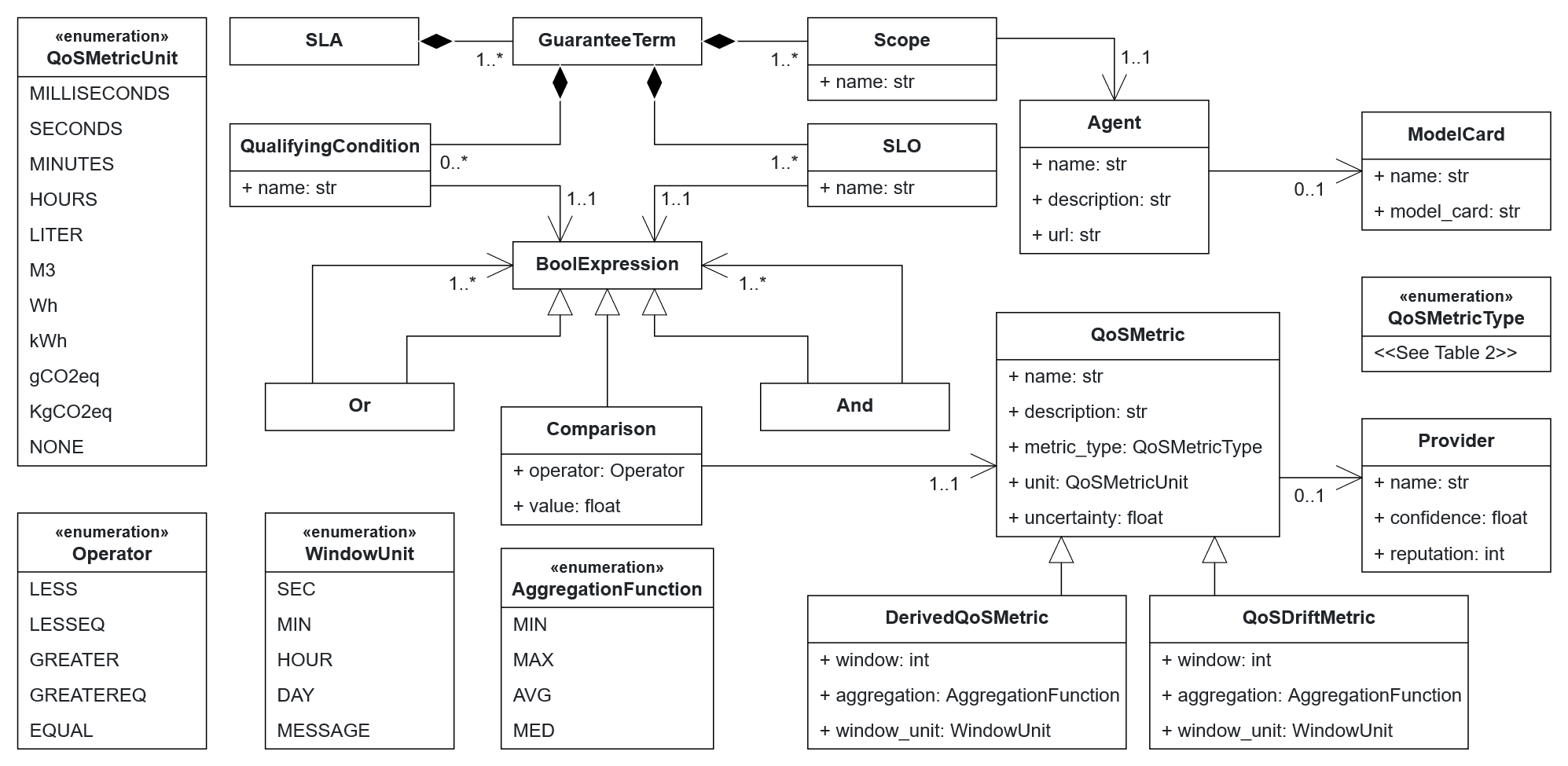}
    \caption{Metamodel of the AgentSLA DSL}
    \label{fig:Metamodel}
\end{figure*}

Figure~\ref{fig:Metamodel} depicts the abstract syntax of AgentSLA in the form of a metamodel (i.e. the schema or grammar of the language expressed using an object-oriented perspective) structuring the language elements, the properties of every element and the possible relationships among them. 
At the root of the metamodel, the \texttt{SLA} class represents the agreement as a set of guarantee terms.
A \texttt{GuaranteeTerm} defines the terms that the service must meet and is composed of scopes, qualifying conditions and Service-Level Objectives (SLOs).
\texttt{Scopes} are identified by a name and relates to the \texttt{Agent} used as scope, itself referring to the potential AI model at use through its Model Card~\cite{mitchell2019model}.
A \texttt{QualifyingCondition} describe the preconditions that must be met to enforce the SLOs in the form of a boolean expression.
An \texttt{SLO} specify a QoS condition that must be assessed and that represents the QoS agreement between service providers and consumers.
This condition is defined as a boolean expression using the \texttt{BoolExpression} class of the metamodel.
Boolean expressions can be conjunctions (\texttt{And}), disjunctions (\texttt{Or}) or comparisons.
A \texttt{Comparison} is defined using a QoS metric (operator left-hand side), a value (operator right-hand side) used as threshold for the metric, and the operator defining the relation between the value and the metric.

The \texttt{QoSMetric} is represented using a name, description, a metric type based on Table~\ref{tab:metrics}, the unit of the metric, and an uncertainty value.
The uncertainty allows addressing the impact of AI models' nondeterministic behavior on the metric measurement and its impact during SLOs evaluation.
Furthermore, \texttt{QoSMetric} refers to their \texttt{Provider} (if known) detailing the confidence in measurement precision and reputation (i.e., similar to GitHub stars) of the provider.
To cover more scenarios, the DSL provides two other types of QoS metrics.
First, \texttt{DerivedQoSMetric} represent the aggregation of multiple measurement in a certain time window, allowing the \texttt{Comparison} to be made with an average or median value for instance.
Thus, it includes a window size, a \texttt{WindowUnit}, and an \texttt{AggregationFunction}.
Second, \texttt{QoSDriftMetric} describe the evolution of a metric across time.
In the context of AI Agent, evolution of the model context can change the overall behavior of the agent.
This can lead to an enhancement or degradation of some QoS metrics.
To detect and assert the stability of the agent, \texttt{QoSDriftMetric} evaluate the difference between two comparable derived QoS metric.
For instance, with a window of ten seconds and using the average aggregation, the \texttt{QoSDriftMetric} will compute the average of the metric between \(t_{-10}\) and \(t_{0}\) to which it will subtract the average between \(t_{-20}\) and \(t_{-10}\).

\subsection{AgentSLA Concrete Syntax}

The concrete syntax of AgentSLA is based on a JSON syntax to allow SLA exchange  and quality information transfer  using existing agent protocols, themselves based on JSON.
A JSON structure is represented using three types of values: objects, arrays and primitive values.
Objects are represented with curly brackets containing key-value pairs using a quoted string for the key, a colon as separator and value representation.
Arrays are represented with square brackets containing values separated by a comma.
Finally, primitives values are represented as usual for programming languages.

To encode SLAs in this format, we defined a set of rules.
\begin{enumerate}
  \item Class instances are represented as objects
  \item Attributes are represented as key-value pairs in the object
  \item Compositions are defined by having the composed class instance as av value of a key-value pair named after its class name
  \item Multiplicity higher than one is managed using arrays
  \item Simple associations are defined using the associated class name as key and the object name attribute as value
\end{enumerate}

In addition, some classes are not direct components of the SLA class (i.e., \texttt{Agent}, \texttt{ModelCard}, \texttt{Provider}, \texttt{QoSMetric} and its subclasses), hence their instances need to be defined in dedicated array in the \texttt{SLA} instance.
Listing~\ref{lst:example} shows an example agreement expressed using the AgentSLA DSL.

\lstdefinestyle{mystyle}{
    keywordstyle=\color{magenta},
    numberstyle=\tiny\color{gray},
    basicstyle=\ttfamily\small,                   
    captionpos=b,                    
    keepspaces=true,
    xleftmargin=10pt,
    numbers=left,                    
    numbersep=5pt
}
\lstset{style=mystyle}
\begin{lstlisting}[float=!htb,label=lst:example,caption=AgentSLA syntax example]           
{
  "GuaranteeTerm":[{
    "Scope":[{
      "name":"Scope 1",
      "Agent": "Agent 1"
    }],
    "QualifyingCondition":[],
    "SLO":[{
      "name": "SLO 1",
      "BoolExpression": {
        "type":"Comparison",
        "QoSMetric": "AVG TTFT",
        "operator": "LESS",
        "value": 1
      }
    }]
  }],
  "DerivedQoSMetric":[{
    "name": "AVG TTFT",
    "description": "description",
    "metric_type": "TTFT",
    "unit": "sec",
    "uncertainty": 0,
    "window": 10,
    "window_unit": "MESSAGE",
    "aggregation": "AVG",
    "Provider": "Provider 1"
  }],
  "Provider": [{
    "name": "Provider 1",
    "confidence": 0.95,
    "reputation": 50
  }],
  "Agent":[{
    "name": "Agent 1",
    "description": "description text",
    "url": "agent_url",
    "ModelCard": "GPT 4o"
  }],
  "ModelCard":[{
    "name": "GPT 4o",
    "model_card": "..."
  }]
}
\end{lstlisting}

In this agreement, we define one guarantee term without qualifying condition, a scope referring to the agent named "Agent 1", and with one SLO named "SLO 1".
The SLO defines one QoS condition expecting the "AVG TTFT" metric to be less than one.
If we look at the \texttt{DerivedQoSMetric} section, we can see the definition of this metric.
The "AVG TTFT" is a derived metric representing the average Time-To-First-Token (TTFT) in seconds across the last ten messages.
Thus, the SLO describe a condition of average TTFT inferior to one second.
The three following sections give additional information about the provider of the metric, the agent and the model used by the agent.
The provider section describes a metric provider called "Provider 1" that has been recommended fifty times and has a confidence value of \(0.95\).
The agent called "Agent 1" is described by a name, description, a URL to reach it and a reference to the model card of the model used in the agent.
This model (called "GPT 4o") is defined by its name and by the textual representation of its model card.

\subsection{Tool support}
To support the definition of SLAs for AI Agents, we provide a Python implementation of AgentSLA available in open-source on GitHub\footnote{Implementation: \url{https://anonymous.4open.science/r/AgentSLA-A813/README.md}}.
This implementation is composed of a validating parser and a set of classes implementing the metamodel.
To implement the metamodel, we used the BESSER Low-Code platform~\cite{alfonso2024building} to model the AgentSLA abstract syntax and generate the corresponding Python class definitions.
The Python classes generated can then be used to instantiate any model conforming to the specified metamodel, allowing its manipulation by other tools.
For instance, this model could be used to generate and deploy the appropriate monitors to enforce the agreement.

In addition, we have implemented a parser validating and transforming the JSON description to AgentSLA model instances.
First, we transform the textual description into a manipulable Python object using Python standard JSON parser.
Then, we traverse the object structure to perform validation checks.
These validation checks ensure: (1) the presence of units when required, (2) the correspondence of these units to the ones defined in the metamodel, (3)  the correspondence of the QoS metric types, operator used, and aggregation function to the ones defined in the metamodel, and (4) that the provider's confidence value bounds to the [0,1] interval.
Finally, if the structure is validated, the model instance is created by instantiating the metamodel classes with the appropriate data.

\section{Discussion}
\label{sec:discussion}

The AgentSLA DSL is a first step towards a more ambitious goal on the establishment of software quality guidelines and quality assurances for AI agents. 
In this section, we discuss the limitations of the proposed approach, as well as the envisioned future work for the evolution of AgentSLA.

\textbf{Validate the quality model.}
When creating the proposed quality model, we aimed to be as complete as possible by making the union of the multiple models existing for AI software~\cite{gezici2022systematic,ali2022systematic} and adding new dimensions relevant in the context of AI agents.
In addition, by providing this model as an extension of the ISO/IEC 25010 standard, we intended to benefit from existing work from the software quality community.
Yet, this quality model has only been partially validated so far in real agentic systems.  The dimensions themselves were presented, discussed and refined with the partners of the MOSAICO project\footnote{More information: \url{https://mosaico-project.eu/}} to better understand their needs and wishes when it comes to integrate AI agents in their new software development projects. But we plan to conduct a more formal empirical validation by applying AgentSLA in new projects  developed by MOSAICO partners or other industrial partners interested in adopting AgentSLA. 

\textbf{Extending the coverage of AgentSLA.}
In designing the AgentSLA DSL, our objective was to provide a relevant set of metrics aligned with the proposed quality model.
However, the topic of AI agents is highly dynamic and new metrics emerge at a high frequency to keep up with the new advance in the field.
One of the limitation of our core set of metrics is the inability for the user to extends the set of metrics.
We envision as future work the evolution of AgentSLA as an extensible DSL allowing users to define their own metrics by specifying how to compute them or where to find a service able to do so.
This way, AgentSLA will be able to cover a wider set of use cases and have the ability to enforce use-case specific SLOs. 

\textbf{Graphical notation.}
Language syntax preferences often vary among users, depending largely on their technical background. 
For example, users with limited technical expertise typically favor more graphical representations. 
To accommodate this diversity, we intend to extend AgentSLA with additional concrete syntaxes, including a graphical notation and a conversational interface.

\textbf{Integration with AI agents ecosystem.}
AgentSLA focus on the elicitation of SLAs and their guarantee terms.
During the definition of its syntax, we used JSON as a base due to its major use in protocols, especially the A2A Protocol that is gaining traction currently.
However, agents receiving the SLA will ignore it as it is not part of the standard.
We propose as future work, the inclusion of the SLA as part of the A2A protocol as well as providing support for the discussion between the agent and the client as part of the Agent Development Kit (ADK)\footnote{\url{https://google.github.io/adk-docs/}} framework.

\textbf{AgentSLA application scenarios.}
The models of Service-Level Agreements resulting from the use of AgentSLA allow for the automatic processing of the specified guarantee terms, which can benefit multiple scenarios.
The first use-case envisioned is the automatic selection of AI agents based on their expected quality of services.
Two similar agents deployed on different platforms or regions (e.g., local vs cloud, Europe vs US) are likely to have different consumption, time behavior or interoperability with local tools.
A second scenario would be the automatic generation, deployment and/or connection to monitors, allowing the evaluation of the specified metrics.
Furthermore, coupled to the first use-case, a possible extension would be the generation of an orchestrator enforcing the agreement by dynamically changing agents when the guarantee terms are not met.

In these use-cases, the model described by AgentSLA is not descriptive anymore but prescriptive. Information previously dedicated to human readers such as the model card or metric provider become criteria that can be leveraged to select agents based on their model card or metric providers based on their reputation and confidence on the quality of the metrics provided

\section{Conclusion}
\label{sec:conclusion}

In this paper, we have presented an approach to specify Service Level Agreements for AI Agents.
In this sense, we proposed (1) a quality model extending the ISO/IEC 25010 standard that propose a set of new quality characteristics relevant in the context of AI agents, (2) a set of core quality metrics aligned with the newly proposed quality characteristics, and (3) a metamodel and concrete syntax allowing the description of Service Level Agreements based on the quality model and metrics previously mentioned, while being compatible with existing protocols for AI agents communication.
On top of these contributions, we provide a Python implementation of the AgentSLA DSL in the form of a validating parser instantiating a model of the agreement conforming to the specified metamodel.

As future work, we plan to advance on some of the discussion points mentioned before. Additionally, we plan to validate the quality model and implement additional tool support for the AgentSLA DSL to generate and/or connect to monitors to automate the monitoring and enforcement of the SLA.

\bibliographystyle{plain}
\bibliography{litterature}

\appendix
\pagenumbering{gobble}
\begin{table*}[!h]
\section*{Appendices}
\centering
\small
\renewcommand{\arraystretch}{1.3}
\hspace{-15pt}
\begin{tabular}{m{35mm}m{45mm}m{85mm}}
\multicolumn{1}{c}{\textbf{Quality Metric}} & \multicolumn{1}{c}{\textbf{Parent in the QM}}                                & \multicolumn{1}{c}{\textbf{Definition}}                                                                                                \\ \hline
Precision                                   & Functional completeness                                                      & True positive over all predicted positive                                                                \\ \hline
Recall                                      & Functional completeness                                                      & True positive over all actual positive                                                                    \\ \hline
Accuracy                                    & Functional correctness                                                       & Percentage of correct prediction                                                                                                       \\ \hline
AUC (Area Under Curve)                      & Functional correctness                                                       & Probability that the model, if given a randomly chosen positive and negative example, will rank the positive higher than the negative. \\ \hline
F1 Score                                    & Functional completeness                                                      & Harmonic mean of accuracy and recall                                                                                                   \\ \hline
XAccuDiff (Cross-validation accuracy difference) & Functional appropriateness                                                   & Accuracy difference between the train and test sets                                                                             \\ \hline
PMV (Perturbed Model Validation)                  & Functional appropriateness                                                   & Accuracy decrease rate between a model and the same model trained with noise in the training data~\cite{zhang2019perturbed}            \\ \hline
TrainingTime                                & Functional appropriateness                                                   & Training time used as complexity proxy~\cite{kirk2014thoughtful}                                                                       \\ \hline
PointwiseRobustness                        & User error protection                                                        & Minimum input change affecting model prediction~\cite{bastani2016measuring}                                                   \\ \hline
AdversarialFrequency                       & User error protection                                                        & Input change impact frequency~\cite{bastani2016measuring}                                                                              \\ \hline
AdversarialSeverity                        & Fault-tolerance                                                              & Distance between an input and its nearest adversarial example~\cite{bastani2016measuring}                                              \\ \hline
AdversarialDistance                        & Fault-tolerance                                                              & AdversarialSeverity on a training input~\cite{tjeng2017evaluating}                                       \\ \hline
TTFT (Time-To-First-Token)                  & Time-behavior                                                                & Time between the request and the generation of the first token~\cite{thangarajah2025slaawareness}                                      \\ \hline
E2E (End-to-end response time)              & Time-behavior                                                                & Time elapsed between request and end result                                                                          \\ \hline
TrainingTime                               & Time-behavior                                                                & Training time                                                                                                                          \\ \hline
Bias                                        & Fairness                                                                     & Ratio of successful bias tests passed using LangBiTe~\cite{morales2024dsl}                                                           \\ \hline
Racism                                      & Fairness                                                                     & Ratio for racism tests~\cite{morales2024dsl}                                                                                           \\ \hline
Sexism                                      & Fairness                                                                     & Ratio for sexism tests~\cite{morales2024dsl}                                                                                           \\ \hline
Ageism                                      & Fairness                                                                     & Ratio for ageism tests~\cite{morales2024dsl}                                                                                           \\ \hline
Religious                                   & Fairness                                                                     & Ratio for religious bias tests~\cite{morales2024dsl}                                                                                   \\ \hline
Political                                   & Fairness                                                                     & Ratio for political bias tests~\cite{morales2024dsl}                                                                                   \\ \hline
Xenophobia                                  & Fairness                                                                     & Ratio for xenophobia tests~\cite{morales2024dsl}                                                                                       \\ \hline
SHAP                                        & Interpretability                                                             & SHAP estimation error~\cite{lundberg2017unified}                                                                                       \\ \hline
LIME                                        & Interpretability                                                             & Comparison to interpretable local surrogates~\cite{ribeiro2016should}                                                \\ \hline
EnergyConsumption                           & \begin{tabular}[c]{@{}l@{}}Training impact, \\ Inference impact\end{tabular} & Estimated energy consumption                                                                                                           \\ \hline
WaterConsumption                            & \begin{tabular}[c]{@{}l@{}}Training impact, \\ Inference impact\end{tabular} & Estimated water consumption                                                                                                            \\ \hline
CarbonEmissions                             & \begin{tabular}[c]{@{}l@{}}Training impact, \\ Inference impact\end{tabular} & Estimated carbon emissions                                                                                                             \\ \hline
CarbonOffset                                & Mitigation                                                                   & Percentage of carbon emissions offset by buying Carbon offset credit                                                                   \\ \hline
OutputSize                                  & Conciseness                                                                  & Length of the generated output                                                                                                         \\ \hline
A2A                                         & Interoperability                                                             & The agent can communicate using A2A (0 if no, else 1)                                                                                  \\ \hline
MCP                                         & Interoperability                                                             & The agent can connect to tools via MCP (0 or 1)                                                                               \\ \hline
OversightLevel                             & Autonomy                                                                     & Level of human oversight as defined by Cihon et al.~\cite{cihon2025measuring}                                                          \\ \hline
\end{tabular}%
\caption{List of metrics mapped to the proposed quality model}
\label{tab:metrics}
\end{table*}

\end{document}